\documentclass[a4paper,11pt]{article}

\usepackage{epsfig}
\usepackage{latexsym}
\usepackage[latin1]{inputenc}   
\usepackage{epsfig}

\newcommand{\MSG}[1]{}

\def\@begintheorem#1#2{\par\bgroup{\sc #1\ #2. }\it\ignorespaces}
\def\@opargbegintheorem#1#2#3{\par\bgroup{\sc #1\ #2\ (#3). }\it\ignorespaces}
\def\@endtheorem{\egroup}

\newtheorem{theorem}{Theorem}[section]
\newtheorem{definition}{Definition}[section]

\newtheorem{example}[theorem]{Example}
 
\def\squareforqed{\hbox{\rlap{$\sqcap$}$\sqcup$}}
\def\qed{\ifmmode\squareforqed\else{\unskip\nobreak\hfil
\penalty50\hskip1em\null\nobreak\hfil\squareforqed
\parfillskip=0pt\finalhyphendemerits=0\endgraf}\fi}

\newcommand{\PLAINLISTING}[1]{\begin{description} {\setlength{\itemsep}{-0.3em}{#1}}\end{description}}
\newcommand{\LISTING}[1]{\begin{enumerate}{\setlength{\itemsep}{0.1em}{#1}}\end
{enumerate}}
\newlength{\LI}
\newcommand{\LLx}{\item\hspace*{0\LI}}
\newcommand{\LLxx}{\item\hspace*{1\LI}}
\newcommand{\LLxxx}{\item\hspace*{2\LI}}
\newcommand{\LLxxxx}{\item\hspace*{3\LI}}

\newcommand{\FIG}[1]{\begin{center}
  \mbox{\epsfclipoff\epsffile{#1.eps}}
  \end{center}}

\def\Z{{\rm Z\kern-0.37em Z}}
\def\N{{\rm I\kern-0.25em N}}
\def\CO{{\rm \kern 0.2em\vrule height 0.67em width 0.05e1 depth -0.02em
\kern-0.33em C}}
\def\R{{\rm I\kern-0.25em R }}
\def\B{{\rm I\kern-0.25em B }}
\def\Q{{\rm I\kern-0.55em Q }}
\def\K{{\rm I\kern-0.25em K }}         
\newcommand{\Rpos}{\R_{\geq 0}} 

\newcommand{\Auto}{\mathcal{A}}
\newcommand{\St}{\mathcal{S}}

\newcommand{\Act}{\mathcal{L}}
\newcommand{\trans}{T}

\newcommand{\ini}{\alpha}
\newcommand{\val}{\beta}

\newcommand{\PM}{{\bf P}}
\newcommand{\QM}{{\bf Q}}
\newcommand{\RM}{{\bf R}}

\newcommand{\EM}{{\bf M}}
\newcommand{\Ea}{{\bf M}_a}
\newcommand{\Eoytn}{{\bf M}[\Phi_1 \wedge \neg \Phi_2 ,\Phi_1 \wedge \neg \Phi_2]}
\newcommand{\Eoyty}{{\bf M}[\Phi_1 \wedge \neg \Phi_2 ,\Phi_2]}
\newcommand{\Noytn}{{\bf N}[\Phi_1 \wedge \neg \Phi_2 ,\Phi_1 \wedge \neg \Phi_2]}
\newcommand{\Eontn}{{\bf M}[\Phi_1 \wedge \neg \Phi_2 ,\neg \Phi_1 \wedge \neg \Phi_2]}

\newcommand{\Eax}{{\bf M}_{a_i}}

\newcommand{\av}{{\bf a}}
\newcommand{\xv}{{\bf x}}
\newcommand{\bv}{{\bf b}}

\newcommand{\wv}{{\bf w}}
\newcommand{\vv}{{\bf v}}
\newcommand{\uv}{{\bf u}}
\newcommand{\dvs}{{\bf d}_{seq}}

\newcommand{\I}{{\bf I}}
\newcommand{\Id}{{\bf I}}
\newcommand{\Ze}{{\bf 0}}
\newcommand{\ei}{{\bf e}_i}
\newcommand{\id}{{\bf e}^T}
\newcommand{\idn}{{\bf e}}
 
\newcommand{\Rel}{\mathcal{R}}
\newcommand{\CC}{\St / \Rel}
\newcommand{\CCi}{\St / \Rel_i}
\newcommand{\CCm}{\St / \sim  }
\newcommand{\Cl}{C}
\newcommand{\Clz}{\mathcal{C}_{\mathcal{R}_0}}
\newcommand{\Clf}{\mathcal{C}_{\mathcal{R}_1}}
\newcommand{\Clse}{\mathcal{C}_{\mathcal{R}_2}}

\newcommand{\hm}{\ \widehat{\cdot} \ }
\newcommand{\hp}{\ \widehat{+} \ }
\newcommand{\zero}{{\rm I\kern-0.4em 0 }}
\newcommand{\zeros}{{\rm I\kern-0.2em 0 }}
\newcommand{\one}{{\rm I\kern-0.55em 1 }}
\newcommand{\aufg}{\Phi_1 \ AU_{\bowtie p}^t \ \Phi_2}
\newcommand{\ufg}{\Phi_1 \ U_{\bowtie p}^t \ \Phi_2}
\newcommand{\ufgn}{\Phi_1 \ U_{\bowtie p}^n \ \Phi_2}
\newcommand{\ufgi}{\Phi_1 \ U_{\bowtie p}^{\infty} \ \Phi_2}
\newcommand{\eufg}{\Phi_1 \ EU_{\bowtie p}^t \ \Phi_2}

\begin{document}

\title{Model Checking for a Class of Weighted Automata}

\author{
\mbox{
\begin{minipage}{8cm} \normalsize
\begin{center}
Peter Buchholz \\ \small
  Fakult{\"a}t f{\"u}r Informatik \\ TU Dresden \\ D-01062 Dresden, Germany\\
  p.buchholz@inf.tu-dresden.de
\end{center}
\end{minipage}     
\begin{minipage}{8cm} \normalsize
\begin{center}
Peter Kemper \\ \small
  Informatik IV \\ Universit\"{a}t Dortmund \\ D-44221 Dortmund, Germany\\
  kemper@ls4.cs.uni-dortmund.de
\end{center}
\end{minipage}
}}
\date{}

\maketitle

\begin{abstract}
A large number of different model checking approaches has been proposed
during the last decade. The different approaches are applicable to different model
types including untimed, timed, probabilistic and stochastic models. This paper
presents a new framework for model checking techniques which includes some of
the known approaches, but enlarges the class of models for which model
checking can be applied to the general class of weighted automata.
The approach allows an easy adaption of model checking to models which have not been
considered yet for this purpose. Examples for those new model types for
which model checking can be applied are max/plus or min/plus automata which are well
established models to describe different forms of dynamic systems and
optimization problems. In this context, model
checking can be used to verify temporal or quantitative
properties of a system.
The paper first presents briefly our class of weighted automata, as a very general
model type. Then Valued Computational Tree Logic (CTL\$) is introduced as
a natural extension of the well known branching time logic CTL. Afterwards, algorithms to
check a weighted automaton according to a CTL\$ formula are
presented. 
As a last result, a bisimulation is presented for weighted automata and for CTL\$. 
\end{abstract}

\section{Introduction}

Model checking of finite state systems is an established approach for the
automatic or semi-automatic analysis of dynamic systems from different
application areas. The basic model checking approaches have been proposed for
untimed models and allow one to check the functional correctness of systems. The
general idea of this kind of model checking is to determine the set of states
of a finite state automaton which satisfies a formula of a temporal
logic. 
Common examples of modal logics to express formulas are {\em Linear Time Logic} (LTL) or {\em Computational Tree Logic} (CTL).
For both logics, efficient analysis
algorithms exist that allow the handling of extremely large automata. Nowadays,
several software tools are available that include model checking algorithms, 
allow the automatic analysis of dynamic systems and have been applied to
practical examples from different application areas like hardware verification or software
engineering. An enormous number of
papers on model checking and related topics exists, for relatively recent
surveys we refer to \cite{ClWi96,ClKu96} and \cite{ClGP99} as a textbook.

For several application areas, the proof of functional correctness is not
sufficient to assure the correct behavior of a system. For instance, in 
real-time systems, it has to be assured that a function of
a reactive system performs correctly and takes place in a given time interval. For other systems, we
may tolerate some erroneous behavior if it occurs only with a sufficiently small
probability. In this and similar situations, a basic proof of correctness is not
sufficient. Consequently, model checking approaches have been extended to
handle also timed, probabilistic and stochastic systems. In \cite{HaJo94}, an
extended version of the temporal logic CTL is presented that is denoted as
{\em Probabilistic Real Time Computational Tree Logic} (PCTL). This logic
allows the definition of properties which state that something will happen with a
given probability in a fixed time interval. The logic is interpreted over
finite {\em Discrete Time Markov Chains} (DTMCs). 
The timing is defined by the number of transitions that occur and
probabilities are defined by the transition probabilities of the DTMC. PCTL is
a useful logic to express requirements for real time systems with constant
delays. Other model checking approaches analyze
different forms of timed automata \cite{AlDi94} that are possibly augmented
by different timing models \cite{BrBD02}. 

The mentioned approaches for model checking are all similar but differ
in various details. In particular, the different logics are all
interpreted over an appropriate automata model. The automata models
used in the mentioned approaches are untimed automata for standard
model checking, probabilistic automata for timed and probabilistic
model checking, stochastic automata for stochastic model checking and
different forms of timed automata.
By considering the wide area of finite state automata, one can notice
that apart from these automata types other models have been
proposed and applied successfully in different application areas.
Examples are min/plus, max/plus, or min/max automata that have been used for
the analysis of real time systems \cite{BaGS99}, communication system
\cite{BaHo00}, and discrete event systems \cite{BCOQ92,Gaub95}. Furthermore,
similar models have been applied for natural language processing
\cite{MoPR96} or image compression \cite{JiLV00}. It is quite natural
and for most of the mentioned applications also very useful to extend
model checking approaches to all these types of automata.  Since the class of
weighted automata provides in some sense a superset of different automata
types, which includes different forms of probabilistic automata and also untimed
automata, one may strive for a general framework of model checking
which can be applied to a wide variety of different types of weighted automata
without defining a new approach for each type. 
Such a framework is of theoretical interest to get a better understanding of
modelchecking and to get a common ground for model checking in various
application areas. From a methodological point of view, it gives direct access
to model checking techniques for various types of automata that do not
profit from these techniques yet. Finally, it supports tool development: in an
object oriented setting, implementation of a specific model checker can
inherit basic techniques from a more general class that implements techniques
valid for the whole framework.

Weighted automata \cite{Eile74,KuSa86} are a well known class of automata where
transitions are labeled with labels from a finite alphabet and,
additionally, receive weights or costs that are elements of
some semiring. A key observation is that the algebraic structure of a semiring is sufficient to define 
modelchecking for weighted automata. The advantage is that by selecting appropriate semirings,
one obtains different types of automata that include most of the above
mentioned types. This general type of automata is suitable to define
a bisimulation as we did in \cite{Buch01,BuKe03}.
In \cite{BuKe01}, the process algebra GPA
has been introduced for the specification of models in a compositional
way such that the underlying semantic model is a weighted automaton 
in the case of a finite set of states.

In this paper, we develop a model checking approach for weighted automata.
The approach allows us to check formulas of the newly defined
logic {\em Valued Computational Tree Logic} (CTL\$) over a weighted automaton.
Algorithms for model checking are developed and it will be
shown that by an appropriate definition of the semiring used for the definition
of transitions weights, we naturally define model checking approaches for
different model types without developing new approaches in each case. The
special cases include untimed, probabilistic, min/plus, max/plus, and min/max automata such
that known model checking approaches are covered and new approaches are introduced in the case of min/plus,
max/plus, and min/max automata. By the use of
other semirings for transition weights, the proposed
approach applies to a  wide class of automata models. In so far, we develop some
form of a generic approach for model checking that is applicable to
other model classes and that includes algorithms to perform model
checking.
 
The structure of the paper is as follows. In the next section, we present the
automata model that is considered in this paper. Afterwards, we define CTL\$, a
logic for automata with transition weights that is
an extension of the well known branching time logic CTL for untimed automata.
The following section introduces algorithms to check a CTL\$ formula according
to an automaton with transition weights. We consider algorithms with explicit state representations
for clarity at this point. A treatment by symbolic representations like {\em
multi terminal binary decision diagrams} (MTBDDs) is feasible but not in the
focus of this paper. In Section \ref{sec:bisimulation},
bisimulation is briefly defined for automata with transition weights and it is
proved that bisimilar automata are indistinguishable under CTL\$
formulas. Afterwards, in Section \ref{sec:examples}, we present several examples
of concrete realizations of weighted automata.
The paper ends with the conclusions.

\section{Weighted Automata}
\label{automata-model}

To present our general automata model, we first introduce semirings that are needed to
define labels for transitions. Afterwards the automata model is defined.

\begin{definition}
\label{def:semiring}
A semiring $(\K,\hp ,\hm, \zero , \one )$ is a set $\K$ with binary operations $\hp$ and
$\hm$ defined on $\K$ such that the following axioms are satisfied:
\begin{enumerate}
\item $\hp$, $\hm$ are associative,
\item $\hp$ is commutative,
\item right and left distributive laws hold for $\hp$ and $\hm$,
\item $\zero$ and $\one$ are the additive and multiplicative identities with $\zero \neq \one$,
\item  $k \hm \zero = \zero \hm k = \zero$ holds for all $k \in \K$.
\end{enumerate}
\end{definition}

Semirings can show specific properties like idempotency, commutativity,
or being ordered; properties that we formally define as follows.
\begin{definition}
\label{def:orderedsemiring}

A semiring is ordered with some transitive ordering $\le$, if  $a \le b$ or $b \le a$ for all
$a, b \in \K$.

An ordered semiring preserves the order if for all $a,b,c \in \K$:
\[
a \le b \ \Rightarrow a \hp c \le b \hp c \ , \ a \hm c \le b \hm c \mbox{ and } c \hm a \le c \hm b . 
\]

A semiring is commutative if multiplication is commutative. 

It is idempotent if addition is idempotent.

It is closed if infinite addition is defined and behaves like finite addition.
\end{definition}

Furthermore, we define $a < b$ if $a \le b$ and $a \neq b$. The supremum $\sup(a,b)$ of $a,b \in \K$ is $a$ if $a > b$ and $b$ otherwise, the infimum $\inf(a,b)$ is $a$ if $a<b$ and $b$ otherwise.
To make the notation simpler, we use sometimes $\K$ for the whole
semiring and $a b$ is used for $a \hm b$.

The well known Boolean semiring $(\B,\vee,\wedge,0,1)$ is order preserving,
commutative, idempotent, and closed whereas $(\Rpos ,+,-,0,1)$ is order
preserving and commutative, but not idempotent and not closed. The semirings
$(\Rpos \cup\{-\infty\},\max,+,-\infty,0)$ and $(\Rpos
\cup\{\infty\},\min,+,\infty,0)$ are order preserving, idempotent, and
commutative, but not
closed. However, they are closed if $\infty$ or respectively $-\infty$ are added.  

\begin{definition}
\label{def:automaton}
A finite weighted automaton over semiring $\K$ and over a finite
alphabet $\Act$ is a $4$ tuple $\Auto=(\St, \ini, \trans, \val)$, where
\begin{enumerate}
\item $\St = \{0,\ldots,n-1\}$ is the finite state space,
\item $\ini : \St \rightarrow \K$ is the initial weight function,
\item $\trans : \St \times \Act \times \St \rightarrow \K$ is the transition
function,
\item $\val : \St \rightarrow \K$ is the final weight function.
\end{enumerate}
\end{definition}

The transition function $\trans$ computes a transition weight for each label and each
pair of states. Independently of the used semiring,
$\trans(x,a,y)=\zero$ implies that no $a$-labeled transition between state
$x$ and state $y$ exists. However, $\zero$ is defined differently in
different semirings. Observe that the definition assures that between
two states at most one transition exists that is labeled with a fixed label $a$. 
For some automata models, initial and final weight 
functions are not needed and hence usually not defined. If this is the case, the functions may be
substituted by constant $\one$, which is the neutral element according
to multiplication; this allows a uniform formal treatment.

\begin{figure}[ht]
  \FIG{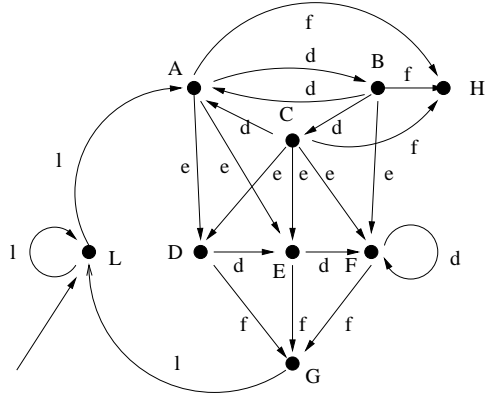} 
  \caption{Example automaton, a driving test model.}
  \label{ex:drivingschool}
\end{figure} 

\begin{example} \label{ex:first}
  We consider a simple model of a driving test to illustrate the
  concept, Fig.~\ref{ex:drivingschool} gives an automaton with $\St
  \{ A, B, \ldots, H, L\}$ and  $\Act = \{l,d,f,e\}$. Let $L$ be the
  initial state, so we define the initialization function $\ini(s) =\one$ if $s = L$ and $\zero$ otherwise. When a student starts,
  he/she takes a couple of lessons, which are transitions with label
  $l$ (\underline{l}esson). After at least one lesson, the student may
  be confident to start the test (state $A$) and drive around with the
  examinator (actions with labels $d$ for \underline{d}rive). While
  driving the examinator may decide to finish (actions with label $f$)
  and to assign the desired driver's license in state $H$ (\underline
  {H}ooray). Alternatively, the student may make some errors (actions
  with label $e$ for \underline{e}rror) that lead him/her to less
  hopeful situations $D$, $E$, or $F$ where after some further driving
  the examinator will finally decide to finish (actions with label
  $f$) and to refuse the license. This yields state $G$. Hence, the
  poor student can only return to $L$ and take some more lessons.
  State $H$ is the desired result, so we define $\val(s)=\one$ if $s=H$ and $\zero$ otherwise.  We will consider this example with
  different semirings and different transition functions (assignment
  of weights), e.g., the Boolean semiring $(\B,\vee,\wedge,0,1)$ is
  useful to ask for existence of paths that lead to a driver's license
  (state $H$), or whether all paths lead to this state, i.e., if success
  is guaranteed. For the Boolean semiring, function $T$ is defined by
  assigning $\one=1$ to all arcs present in
  Fig.~\ref{ex:drivingschool}. $([0,1],+,\cdot,0,1)$ is useful to
  achieve a probabilistic model, where actions are randomly selected
  and one may ask for the probability to succeed. If one asks for
  trouble, one can use $(\Rpos\cup\{-\infty\},max,+,-\infty,0)$ to
  look for the hardest path to success, and
  $(\Rpos\cup\{\infty\},min,+,\infty,0)$ for the one with minimal
  stress, given that weights indicate how much energy is necessary to
  perform that action.
\end{example}
The class of weighted automata is known for a long
time in automata theory \cite{Eile74}. The concrete realization defined here has
been proposed in \cite{Buch01}. In \cite{BuKe01}, a process algebra is
presented that is based on the above concepts in the sense that its
dynamic behavior yields an automaton with transition
weights; however a term in a process algebra may impose an automaton
with an infinite number of states. Different semirings yield completely
different automata. Before we present some concrete realizations by fixing the
semiring, we consider
general methods to analyze the behavior of automata with transition
weights. The behavior of a weighted automaton considers the weights of paths
between states where a path is described by a finite or infinite sequence of
transition labels.

To analyze the behavior of an automaton over all paths, we present an approach
that is based on vector-matrix computations because this is a
convenient approach to compute these results. Since we consider automata over 
finite state spaces and finite sets of transition labels, each automaton can
be  described by sets of $\K^{n \times n}$ matrices
and $\K^n$ vectors. Thus, we define for each $a \in \Act$ a matrix $\Ea$ with
$\Ea(x,y)=\trans(x,a,y)$ and $\EM = \widehat{\sum}_{a \in \Act}\Ea$ as a matrix
that collects all weights independently of the labels. Furthermore, we define
a row vector $\av$ with
$\av(x)=\ini(x)$ and  a column vector $\bv$ with $\bv(x)=\val(x)$.
To complete the notation let, $\I$ be the $n \times n$
identity matrix over semiring $\K$, let $\ei \in \K^n$ be a row vector with $\one$ in
position $i$ ($0 \le i < n$) and $\zero$ elsewhere and let $\id$ be a column
vector where all elements equal to $\one$. It is straightforward to define
matrix sum and product using the operations of the semiring instead of the
usual multiplication and addition. 

We start with the analysis of paths and introduce some
notations first.  We use $x$, $y$, $z$ for states and $i$, $j$, $k$
for running indices in sums or products. A path of automaton $\Auto$
is defined as a sequence of states and transitions starting in a state
$x \in \St$ with $\ini(x) \neq \zero$. In automata theory paths may be
defined by sequences of states or transitions or both. We use here a
definition that observes transitions via their labels and states.
However, the approach can be easily restricted to observe only states
or only transitions. Let $\pi$ be a path, $\pi^s$ the sequence of
states in the path and $\pi^t$ the transition labels. We denote by
$\pi^s_i \in \St$ ($i = 0, 1, 2, \ldots$) the $i$-th state in the path
and by $\pi^t_j \in Act$ ($j = 1, 2, \ldots$) the $j$-th transition
label.   Thus,
$\pi=(\pi_0^s\pi_1^t\pi_1^s\dots)$ is a path of automaton $\Auto$ if
$\trans(\pi^s_i, \pi^t_{i+1}, \pi^s_{i+1}) \neq \zero$.  A path might
be of infinite or finite length. In the finite case, index $i$ runs
from $0$ to $| \pi |$ where $| \pi |$ is the length of the path, i.e.,
the largest index $i$ in the path. Let $\sigma$ be the set of paths of
automaton $\Auto$, $\sigma^n$ ($\sigma^{\le n}$) the set of paths of
length $n$ ($\le n$) and $\sigma^n_x$ ($\sigma^{\le n}_x$) the set of
paths of length $n$ ($\le n$) that start in state $x$. For each finite path,
we can compute the weights as (ce = \underline{c}osts \underline{e}ach)
\begin{equation}
\label{costs-0}
ce(\pi) = 
\ini(\pi^s_0) \hm \widehat{\prod}_{i=0}^{| \pi |} \trans(\pi^s_i, \pi^t_{i+1},
\pi^s_{i+1}) \hm \val(\pi^s_{|\pi|})
\end{equation}
where $\widehat{\prod}_{i=1}^N a_i = a_1 \hm \ldots \hm a_N$ and the case of finite $N$ might be
extended to $N = \infty$, if the semiring is appropriately chosen such
that the infinite product can be computed.

If we focus on observing the behavior of an automaton by considering a
sequence of labels $seq=a_1,\ldots,a_m$ with $a_i \in \Act$ for a path $\pi$,
then one does not want to distinguish among paths $\pi$ and $\pi'$ that
produce the same sequence $seq$.
The weights summed over all paths with labeling
$seq$ 
starting in state $x$ are given by (ca = \underline{c}osts \underline{a}ll)
\begin{equation}
\label{costs-1}
ca_x(seq) = \av(x) \hm (\widehat{\prod}_{i=1}^m \Eax ) \hm {\bv} \ ,
\end{equation}
and the weights of all paths of length $m$ with an arbitrary labeling is
computed as
\begin{equation}
ca_x(*^m) = \av(x) \hm  \EM^m \hm {\bv}^T \ .
\end{equation}
The above computation of weights assumes that a specific initial state is
known. Alternatively, one can consider the case that vector $\av$ defines the
weights of initial states. The weights of paths are defined then as
\begin{equation}
\label{costs-2}
ca(seq) = \av \hm (\widehat{\prod}_{i=1}^m \Eax ) \hm {\bv} \mbox{ and }
ca(*^m) = \av \hm \EM^m \hm {\bv} .
\end{equation}
Apart from the weights of paths, we consider possible terminating states and the
weights of reaching those states. These values are described by a row vector
\begin{equation}
\label{costs-3}
\dvs = \av \hm (\widehat{\prod}_{i=1}^m \Eax ) \ ,
\end{equation}
such that $ca(seq)=\dvs \hm \bv$.
 
\section{Valued Computational Tree Logic}

The usual way of describing dynamic properties of a system are
temporal logics which exist in various forms. Very popular is the
branching time logic CTL \cite{ClES86}. CTL formulas are interpreted
over labeled transition systems and efficient algorithms for model
checking finite systems exist \cite{BCMDH92} and have been implemented
in software tools \cite{ClPS93}. CTL allows us to check properties of
paths of an automaton where an all- or existence-quantifier has to precede
any path quantifier. Since CTL is defined for
transition systems where transitions are not quantified, it cannot be
used to derive properties that hold with a certain probability or
hold for a specified time. To express such probabilities, the logic has
to be extended as done by several authors. The logic RTCTL
is described in \cite{EMSS92} as an extension of CTL. RTCTL, in
contrast to CTL, allows reasoning about times. Thus, it can be
expressed that a property will become true within 50 time units or that
a property holds for 20 time units. Time is discrete in this model and
one transition takes exactly one time step. In \cite{HaJo94}, the logic
PCTL is introduced that can be used to describe properties that
hold for some time (or after some time) and hold with at least a given
probability. Thus, this logic extends RTCTL with respect to
probabilities. Formulas of PCTL are interpreted over discrete time
Markov chains (DTMCs) and  the model checking
problem for PCTL is polytime decidable \cite{BeSl98}. In
this model, time is also
discrete and one transition lasts one time step. 

In this paper, we extend CTL by defining a logic for weighted automata.
This approach is more general than the
previous extensions of CTL because it can be applied to a large number of
models by defining an appropriate semiring structure for quantifying
transition labels. Since our automata model contains transition labels we
extend our logic by propositions that allow us to reason over labeled
transitions as it is done in Hennesy-Milner logic
\cite{HeMi85,Miln89}. In this respect,
{\em Valued Computational Tree Logic} (CTL\$) might not be the natural
name for the logic. However, since CTL is included in the logic CTL\$ as a special
case of automata over the Boolean semiring, we choose this name. 
We will show later that the approach includes probabilistic systems, although the
presented logic is in these cases not completely equivalent to the different
logics proposed for the models mentioned above. We will come back to this point in Section
\ref{sec:examples} where we present concrete realizations of our model. Here, we first
define basic CTL\$ formulas, introduce informally the semantics of a formula, and
define some derived expressions afterwards.

\begin{definition}
\label{def:gctl} For a given set of atomic propositions, the syntax of
a CTL\$ formula for a semiring $\K$ is defined  inductively as
follows:
\begin{itemize}
\item An atomic state proposition $\Phi$ is a CTL\$ formula,
\item if $\Phi_1$ and $\Phi_2$ are CTL\$ formulas, then $\neg \Phi_1$ and $\Phi_1 \vee \Phi_2$ are CTL\$
formulas,  
\item if $\Phi$ is a CTL\$ formula and $p \in \K$, then 
$[a]_{\bowtie p}. \Phi$ is a CTL\$ formula, and
\item if $\Phi_1$ and $\Phi_2$ are CTL\$ formulas, $t$ is a nonnegative integer or
$\infty$ and $p \in \K$, then $\ufg$ and $\aufg$ are CTL\$ formulas
\end{itemize}
where $\bowtie \in \{<, \le, =, \ge, > \}$.
\end{definition}

Formulas of CTL\$ are interpreted over weighted automata.
A necessary condition to interpret a formula for an automaton is that
both use the same semiring $\K$, which will be assumed in the sequel. Atomic
propositions of the kind $\Phi : \St \rightarrow \B$ describe
elementary properties that hold or do not hold in a state $s \in \St$
of an automaton.  The goal of model checking is to compute the set of
states for which a CTL\$ formula $\Phi$ holds. Before we define
formally for which states a formula holds, we present the intuitive
meaning of the formulas, i.e., we describe under which conditions
formula $\Phi$ holds for state $x$.

\begin{itemize}
\item An atomic proposition $\Phi$  is true in $x \in \St$, if the proposition
holds in $x$.
\item $\neg \Phi$ is true in $x$ if $\Phi$ is false in $x$;  $\Phi_1 \vee \Phi_2$ is true in
$x$ if $\Phi_1$ or $\Phi_2$ are true in $x$.
\item $[a]_{\bowtie p}.\Phi$ is true in $x$ if $w \bowtie p$ holds
  where $w$ denotes  the sum of weights of
$a$-labeled transitions that leave $x$ and end in some state where $\Phi$
holds.
\item $\ufg$ is true in $x$, if $w \bowtie p$ holds  where $w$ denotes the total amount of weights
for all paths that 1) start in $x$ and  2)
fulfill $\Phi_1$ until they reach a state where $\Phi_2$ holds, and 3) 
perform at most $t$ steps for condition 2). This operator ignores those paths 
that fail on any of the conditions  1) - 3). 
\item $\aufg$ is true in $x$, if all paths that 1) start in $x$, 2) fulfill
$\Phi_1$ until they reach a state where $\Phi_2$ holds, 3) require for this at
most $t$ steps and for the sum of the weights $w$ of all these paths $w \bowtie
p$ holds. This operator is more strict than the previous one, it requires 
all paths to observe conditions 1) - 3). 
\end{itemize}

We use the notations $x \models \Phi$ if $x$ satisfies formula $\Phi$ and
$\neg x \models \Phi$ if this is not the case. 
The meaning of the first two cases above is obvious. 
For a formal definition of the last three cases, 
we make use of a description by vectors and matrices and introduce
some additional notations first. Let for some matrix $\RM \in \K^{n\times n}$
and two $CTL\$$ formulas $\Phi_1$ and $\Phi_2$,
$\RM[\Phi_1,\Phi_2]\in\K^{n\times n}$ be defined as
\[
\RM[\Phi_1,\Phi_2](x,y)= \left\{
\begin{array}{ll}
\RM(x,y) & \mbox{if } x \models \Phi_1 \mbox{ and } y \models \Phi_2 \\
\zero & \mbox{otherwise} 
\end{array}
\right.
\]
Consequently, matrix $\EM[\Phi_1,\Phi_2]$ ($\Ea[\Phi_1,\Phi_2]$)
contains all transitions (labeled with $a \in \Act$) that start
in a state where $\Phi_1$ holds and end in a state where $\Phi_2$
holds and $\Id[\Phi,\Phi]$ is a matrix that contains $\one$ in the main
diagonal whenever $\Phi$ holds for the corresponding state and all other
elements are $\zero$. Furthermore, let for some vector $\xv$,
$\xv[\Phi]=\xv \Id[\Phi,\Phi]$.
With these notations, we can formally define the meaning of the
presented CTL\$-formulas using vectors and matrices rather than
considering specific paths.

\begin{itemize}
\item $x \models [a]_{\bowtie p}.\Phi$ if and only if $w \bowtie p $ with  $w=\idn_x \Ea \id[\Phi]$.
\item $x \models \ufg$ if and only if  
$w \bowtie p $ with
\[
w = \left\{
\begin{array}{ll}
\av(x) \idn_x \left(\widehat{\sum}_{k = 0}^{t-1}(\Eoytn)^k\right)\hm \Eoyty \hm
\bv[\Phi_2] & \mbox{ if } t > 0\\
\av[\Phi_2](x) \hm \bv[\Phi_2](x) &   \mbox{ if } t = 0
\end{array}
\right.
\]
\item $x \models \aufg$ if and only if $x \models \ufg$ and for all $\pi \in
\sigma_x$ exists some $m \le t$ such that $\pi^s_m \models \Phi_2 \wedge \pi^s_i
\models \Phi_1 \wedge \neg \pi^s_i \models \Phi_2$ for $0 \le i <
m$.
\end{itemize}

If a semiring is ordered, preserves the order by its operations and
 $\zero$ is the infimum of $\K$, then
$a\hp b = \zero$ implies $a = b = \zero$. In that case, we 
can equivalently rewrite the condition on paths $\pi$ for $x \models
\aufg$ by requiring that the sum of weights of paths that contradict the
property is $\zero$. More formally,
\begin{equation}
\label{eq:au_check}
\begin{array}{ll}
\av(x) \hm \idn_x \hm \big((\widehat{\sum}_{k = 0}^{t-1}(\Eoytn)^k) \hm  \Eontn \\ \\
\hspace*{1.8cm}  \hp (\Eoytn)^t\big)\hm \id & = \zero 
\end{array}
\end{equation}
for $t > 0$. For $t = 0$, we have  $x \models \ufg
\Leftrightarrow x \models \aufg$.

CTL\$ contains an all but no existence quantifier. The reason for this
decision is that the existence quantifier can often be described by the general
path quantifier $U$ using $U_{>\zeros}^t$, which indicates for many, but not for
all semirings that a path of length $\le t$ exists that observes the required
properties. For instance, the boolean semiring is a case where
$U_{>\zeros}^t$ is suitable to decide existence of a path.

Another reason for not introducing an existence quantifier for
paths is that in general semirings this quantifier is not indistinguishable
under bisimulation. Thus, bisimilar automata (see Sect. \ref{sec:bisimulation})
still might be distinguished via CTL\$ formulas including path
quantifiers considering single paths and
this is in some sense against the idea of bisimulation and its connection to
logics. 
Note that the quantifier $AU$ does not introduce
problems for order preserving semirings where $\zero$ is the infimum
and these semirings will be considered in the algorithms presented
below. $AU$ is necessary to make CTL\$ equivalent to CTL
if weighted automata are defined over the Boolean semiring. Since
CTL\$ shall not be less expressive than CTL, $AU$ must be included. 

Several other operators can be derived from the basic operators of CTL\$ .
The basic operators $\wedge$ and $\rightarrow$ are derived in the obvious way. By help of negation, one can show
that for path formulas with $\ufg$, not all operators for comparisons $\bowtie$ are essential.
We present the relation for $\ufg$ and omit index $p$, for readability. 
\[
\begin{array}{ll}
\Phi_1 \ U_{<}^t \ \Phi_2 = \neg \left((\Phi_1 \ U_{=}^t \Phi_2) \vee
(\Phi_1 \ U_{>}^t \Phi_2) \right) & \mbox{    }
\Phi_1 \ U_{\le}^t \ \Phi_2 = \neg (\Phi_1 \ U_{>}^t \Phi_2) \\ \\
\Phi_1 \ U_{\ge}^t \ \Phi_2 = (\Phi_1 \ U_{=}^t \Phi_2) \vee
(\Phi_1 \ U_{>}^t \Phi_2) & \mbox{    }
\Phi_1 \ U_{=}^t \ \Phi_2 = (\Phi_1 \ U_{\ge}^t \Phi_2) \wedge \neg
(\Phi_1 \ U_{>}^t \Phi_2)
\end{array}
\]
The last equality shows that we may as well use $\bowtie \in \{>, \ge\}$ to
derive all other relations for $\ufg$. This will be done in the following 
section because $>$ and $\ge$ can be easily checked in the
algorithms. 

%
Similarly, we have the following relation for $[a]_{\bowtie p}$ where
$p$ is again omitted for readability. 
\[
[{a}]_{{<}}  = \neg \left( [{a}]_{{=}} \vee [{a}]_{{>}} \right), \
[{a}]_{{\le}}  = \neg [{a}]_{{>}}, \
[{a}]_{{\ge}} = [{a}]_{{=}} \vee [{a}]_{{>}} \mbox{ and }
[{a}]_{{=}} = [{a}]_{{\ge}} \wedge \neg [{a}]_{{>}}
\]
Again, it is sufficient to consider $\bowtie \in \{>, \ge\}$.
%

The following abbreviations are defined by extending the corresponding CTL\$ formulas. 
\begin{itemize}
\item $AX_{\bowtie p} \ \Phi = true \ AU_{\bowtie p}^1 \Phi$ and
  $UX_{\bowtie p} \Phi = true \ U_{\bowtie p}^1 \Phi$
\item $AF_{\bowtie p}^t \ \Phi = true \ AU_{\bowtie p}^t \Phi$ and $UF_{\bowtie
p}^t \Phi = true  \ U_{\bowtie p}^t \Phi$.
\end{itemize} 

$X$ corresponds to a {\em next} operator. $F$ denotes a {\em finally}
operator. Such operators are common syntactical sugar of modal logics.

\section{Model Checking CTL\$ Formulas}
\label{sec:model-checking} 

To perform model checking in an efficient way, we restrict the
semiring used for transition valuation. We assume that the semiring is
ordered, that the order is preserved by the operations, and that $\zero$ is the
infimum of $\K$. Observe that these conditions are satisfied in most
practically relevant semirings, e.g., in the examples presented below.
To illustrate the point, $(\R,+,\cdot,0,1)$ is not ordered due to the
fact that $a \leq b$ does not imply $ a \cdot c \leq b \cdot c$ if $c
< 0$, but $(\Rpos,+,\cdot,0,1)$ is ordered.  This means that we
prohibit negative weights, which is a common and familiar restriction in
consideration of automata with transition weights. At the end of the
section, we briefly outline when and how model checking can be
performed for more general semirings.

We follow other model checking approaches like \cite{ClES86} and define
inductively over the length of a formula how a formula is checked.
\begin{itemize}
\item $leng(\Phi) = 1$ if $\Phi$ is an atomic proposition,
\item $leng(\neg \Phi) = leng(\Phi) + 1$,
\item $leng(\Phi_1 \vee \Phi_2)= \max (leng(\Phi_1), leng(\Phi_2)) + 1$,
\item 
$leng([a]_{\bowtie p}.\Phi) = leng(\Phi) +
1$ and
\item $leng(\ufg) = leng(\aufg) = \max (leng(\Phi_1), leng(\Phi_2)) + 1$.
\end{itemize}

As in CTL model checking the set of states satisfying a formula of length $l$
is computed after all sets of states that satisfy sub-formulas of length $<l$ are
known. Computation of the sets of states that observe atomic propositions, $\neg
\Phi$, or $\Phi_1 \vee \Phi_2$ is identical to the corresponding computations in
CTL. Thus, the new cases are $[a]_{\bowtie
p}.\Phi$, $\ufg$, and $\aufg$. We describe a procedure for
each of the three formulas that computes for each state whether it observes
the formula or not. We present only cases of  $\bowtie \in
\{>, \ge\}$, 
since the other cases can be derived from these cases as shown above.
Let {\sl marked($x$)} be a variable that is {\sl true} if
$x \models \Phi$ and {\sl false} otherwise. For the presentation of
the algorithms, we use the vector matrix representation of the
automaton which is also well suited for an implementation of the algorithms. 

\paragraph{An algorithm to compute $[a]_{\bowtie p}.\Phi$.}

\PLAINLISTING{
\LLx for (all $x \in \St$) do
\LLxx marked($x$) := false ;
\LLxx sum := $\zero$ ;
\LLxx for (all $y$ with $\Ea(x,y) \neq \zero$) do
\LLxxx if ($y \models \Phi$) then
\LLxxxx sum := sum $\hp \Ea(x,y)$ ;
\LLxxx if (sum $\bowtie p$)
\LLxxxx marked($x$) := true ;
\LLxxxx break ;
}

The inner {\sl for}-loop can be left if sum $\bowtie p$ because due to
our assumptions the value of sum cannot be reduced according to $\le$ or
$<$. 

\paragraph{An algorithm to compute $\ufg$ for $t < \infty$.}

\LISTING{
\LLx for (all $x \in \St$) do
\LLxx if ($x \models \Phi_2$) then
\LLxxx $\wv(x) := \bv(x)$ ;
\LLxxx if ($\av(x) \hm \bv(x) \bowtie p$) then
\LLxxxx marked($x$) := true ;
\LLxxx else
\LLxxxx marked($x$) := false ;
\LLxx else
\LLxxx $\wv(x) := \zero ;$
\LLxxx if ($\neg x \models \Phi_1$) then
\LLxxxx marked($x$) := false ;
\LLxxx else
\LLxxxx marked($x$) := undefined ;
\LLx $\vv := \Eoyty \hm \wv$ ;
\LLx $\uv := \vv$ ;
\LLx for (all $x \in \St$ with marked($x$) = undefined) do
\LLxx if ($\av(x) \hm \uv(x) \bowtie p$) then
\LLxxx marked($x$) := true ;
\LLx $l := 2;$ 
\LLx while ($l \leq t$ and $\exists x \in \St$ with marked($x$)=undefined) do 
\LLxx $\wv := \Eoytn \hm \vv$ ;
\LLxx $\uv := \uv \hp \wv$ ;
\LLxx for (all $x \in \St$ with marked($x$) = undefined) do
\LLxxx if ($\av(x) \hm \uv(x) \bowtie p$) then
\LLxxxx marked($x$) = true ;
\LLxx $l := l + 1$ ;
\LLxx $\vv := \wv$ ;
\LLx for (all $x \in \St$ with marked($x$) = undefined) do
\LLxx marked($x$) := false ;
}
 
Steps 1 through 13 of the algorithm describe the initialization phase, 
several special cases are decided directly. A state
$x$ that satisfies $\Phi_2$ also satisfies $\Phi_1 \ U^{t}_{\geq p} \
\Phi_2$ if $\av(x) \hm \bv(x) \bowtie p$. If ``$\bowtie$'' $=$
``$\geq$'' and $\neg (\av(x) \hm \bv(x) \geq p)$ then $x$ does
not satisfy the formula, because on all paths starting in $x$ $\Phi_2$
immediately holds for the first time and the weights of these paths are too small
such that the whole formula is false. For states where $\Phi_1$ and $\Phi_2$
both do not hold, the formula is false too. In the remaining cases, it is not
clear yet whether the formula holds or not and those states are marked
as {\em undefined} with respect to this formula. Steps 14 through 18
describe the first transition going from a state where $\Phi_1$, but
not $\Phi_2$ holds into a state where $\Phi_2$ holds and check whether the
formula becomes {\em true} by paths of length one. 
In the steps 19 through 27, 
transitions of the
automaton between states where $\Phi_1$ but not $\Phi_2$ holds are
mimicked step by step. In each step $l$, we compute per state $x$,
where only $\Phi_1$ holds, the sum of weights of paths of length $l$ that end in a state 
where $\Phi_2$ holds and that pass through states where only $\Phi_1$ holds. These
weights are collected in vector $\wv$. The weights of all those paths of length
of at  most $l$ are accumulated in vector $\uv$. The iteration stops if $t$ steps have been
computed, which means  that all paths of length $\le t$ have been considered, or
if all states are classified, i.e., no state is marked as undefined.
After leaving the iteration over all paths of length $\le t$, all states
that are still marked undefined do not satisfy the formula because no
appropriate path can be found for them. The procedure eventually stops
for finite $t$.

For $t = \infty$, the situation is different. In principle, we can use
the above procedure, but it cannot be assured whether it stops or
yields to an infinite computation. The crucial point is the
computation of the infinite sum of matrices
\[
\Noytn = \widehat{\sum}_{k=0}^{\infty} (\Eoytn)^k \ .
\] 
The following relation holds if $\Eoytn$ can be reordered to an
upper triangular matrix.
\[
\widehat{\sum}_{k=0}^{\infty} (\Eoytn)^k = 
\widehat{\sum}_{k=0}^{n} (\Eoytn)^k 
\]
The relation is true since $\RM^k=\Ze$ for $k>n$ if  $\RM \in \K^{n,n}$ is an upper triangular matrix. In this case and for $t \ge n$
\[
x \models \ufg \ \Leftrightarrow \ x \models \ufgn
\]
such that the above algorithm for finite $t$ can be applied for the
infinite case as well.

For the general case where $\Eoytn$ cannot be reordered to an upper
triangular form, computation of $\Noytn$ requires that the semiring
$\K$ is closed and the concrete computation depends on the used
semiring. We will give some examples for different semirings below. If
$\Noytn$ is available, then $\ufgi$ can be checked using an extension
of the algorithm for the finite case, where the steps 19 through 29
are substituted by the following steps.
\PLAINLISTING{
\LLx $\uv := \Noytn \hm \vv$ ;
\LLx for (all $x \in \St$ with marked($x$) = undefined) do
\LLxx if ($\av(x) \hm \uv(x) \bowtie p$) then 
\LLxxx marked($x$) := true ;
\LLxx else
\LLxxx marked($x$) := false ;
}

\paragraph{An algorithm to compute $\aufg$.}\ \\
To analyze $x \models \aufg$, first $x \models \ufg$ has to be proved
with the presented algorithm and then (\ref{eq:au_check}) has to be
checked. 
Since we restrict ourselves to an order preserving semiring $\K$ with $\zero$ as its
infimum, the following result holds for $\RM \in \K^{n,n}$, $\av, \bv
\in \K^n$.
\[
\av \widehat{\sum}_{k=0}^{\infty} \RM^k \bv > 0 \ \Leftrightarrow \
\av \widehat{\sum}_{k=0}^{n} \RM^k \bv > 0
\]
The result holds since the existence of a path between two states
implies the existence of a path of length $\le n$ between these
states (remember that $|\St|=n$). Thus (\ref{eq:au_check}) becomes
\[
\begin{array}{ll}
\av(x) \hm \idn_x \hm \left((\widehat{\sum}_{k = 0}^{\min(t-1,n-1)}(\Eoytn)^k) \hm \Eontn\right. \\ \\
\hspace*{1.8cm}\left. \hp (\Eoytn)^{\min(t,n)}\right) \hm \id & = \zero \ .
\end{array}
\]
This relation is checked for $t > 0$ in the following algorithm where we assume
that marked($x$) is {\em true} if $x \models \ufg$ and {\em false}
otherwise. 
\LISTING{
\LLx $\wv := \Eontn \hm \id$ ;
\LLx for (all $x \in \St$ with marked($x$) = true and $\wv(x) > \zero$) do
\LLxx marked($x$) = false ;
\LLx $k := 1$ ;
\LLx while ($k < \min(t,n)$ and $\exists x$ with marked($x$) = true) do
\LLxx $\wv := \Eoytn \hm \wv$ ;
\LLxx for (all $x \in \St$ with marked($x$) = true and $\wv(x) > \zero$) do
\LLxxx marked($x$) = false ;
\LLxx $k := k + 1$ ;
\LLx $\wv := \id[\Phi_1 \wedge \neg \Phi_2]$ ;
\LLx for ($k = 1$ to $\min(t,n)$) do
\LLxx $\wv := \Eoytn \hm \wv$ ;
\LLx for (all $x \in \St$ with marked($x$) = true and $\wv(x) > 0$) do
\LLxx marked($x$) = false ;
}
The procedure checks both conditions on which $\aufg$ may fail
separately and requires a finite effort due to the finite summations.

Evaluation of $\ufg$ and $\aufg$ involves computation of $\RM^t$ and
$\widehat{\sum}_{k=0}^t \RM^k$ as subproblems. In the algorithms given
so far, those subproblems are solved by successive matrix-vector
multiplications, which avoids an explicit computation of $\PM=\RM^t$ and
$\QM=\widehat{\sum}_{k=0}^t \RM^k$. If the space used to represent $\PM$
or $\QM$ is tolerable for an application, those matrices can be
computed with less steps by using iterated squaring if the semiring is
idempotent. Iterated squaring is known for long, e.g., to compute a
transitive closure of graph which corresponds to the boolean semiring. 
To compute $\PM$, we can use a binary representation of
$t=\sum_{j=0}^{l} \delta_j 2^j$ with $l=\lfloor log(t) \rfloor$ and $\delta_j \in \{0,1\}$ 
such that $\PM = \widehat{\prod}_{j=0, \delta_j=1}^{l} \RM^{2^j}$ and
$\RM^{2^j}$ is obtained by computing a sequence $\RM, \RM^{2^1},
\RM^{2^2},\dots,\RM^{2^l}$ with $l$ matrix-matrix multiplications.
Iterated squaring to compute $\PM$ works for semirings in general.
In case of an idempotent semiring, we can use that approach for $\QM$ as
well.
It is straightforward to verify that $(\RM + \Id)^t = \widehat{\sum}_{k=0}^t
\RM^k$ in case of an idempotent semiring. We briefly recall the argument for this
known result. Obviously, the result is true for $t=0$. For $t > 0$, we
first use the induction hypothesis and the idempotency of the
semiring, in this way we have $(\RM + \Id)^t = (\RM + \Id)^{t-1}\hm
(\RM + \Id) = \widehat{\sum}_{k=0}^{t-1}
\RM^k \hm (\RM + \Id) = 
\widehat{\sum}_{k=1}^{t} \RM^k \hp \widehat{\sum}_{k=0}^{t-1} \RM^k = \widehat{\sum}_{k=0}^t \RM^k$. Hence, for idempotent semirings, we can for
instance compute $\widehat{\sum}_{k=0}^{t-1}(\Eoytn)^k$ and 
$ (\Eoytn)^t$ with at most $log(t)$ matrix-matrix
multiplications and additions.

With the presented algorithms, all formulas of CTL\$ can be proved for
the class of semirings that has been defined at the beginning of
this section. The only missing step is the computation of the matrix
$\Noytn$ which has to be realized specificly for each semiring. In the
examples below, we show that computation of this matrix can be done in
most interesting semirings with an effort of $O(n^3)$ or below. If this
is the case, then the effort for checking $\ufgi$ is in $O(n^3)$
whereas the effort for checking $\ufg$ for finite $t$ is in
$O(tn^2)$. In general, the effort grows linear in $t$ and in the length
of the formula and it grows at most cubic in the size of the
automaton.

Checking CTL\$ formulas for more general semirings that are not order
preserving requires some restrictions since otherwise an infinite
summation may not be computable (for instance in case of divergent
sums, non-existence of a fixpoint). Usually, matrix $\Noytn$ cannot be
computed for these semirings such that $\ufg$ can only be checked for
finite $t$. Furthermore, the checking of $\aufg$ often cannot be done
with the presented algorithm. If we restrict the formulas to those
that do not contain $\aufg$ and contain $\ufg$ only for finite $t$,
then the proposed algorithms can still be applied for modelchecking
given that those parts are removed that terminate a loop due to $w \bowtie
p$. All decisions that rely on comparisons $w \bowtie p$ must be
delayed to the end of the procedures since values can change in a
non-monotonous manner.


\section{Bisimulation for Weighted Automata}
\label{sec:bisimulation}

Bisimulation for weighted automata has been introduced in
\cite{Buch01}. In \cite{BuKe01}, it has been shown that bisimulation is a
congruence according to the operations of the process algebra GPA. Here, we
briefly rephrase the definition for bisimulation given in \cite{Buch01,BuKe01} and
prove afterwards that bisimilar states of an automaton are indistinguishable
under CTL\$ formulas.

We consider only equivalence relations as bisimulations. Let $\Rel$ be an equivalence
relation on $\St \times \St$. $\CC$ is the set of equivalence classes of
$\Rel$, $\Cl \in \CC$ is an equivalence class of $\Rel$ and $\Cl[x]$
is the equivalence class to which state $x \in \St$ belongs. If we consider equivalence classes of different equivalence
relations $\Rel_i$, we use $\Cl_{\Rel_i}$ for an equivalence class
from $\CCi$. We define for 
$\Cl \subseteq \St$: $\EM(x,\Cl)=\widehat{\sum}_{y \in \Cl} \EM(x,y)$.

\begin{definition}
\label{def:bisimulation}
An equivalence relation $\Rel$ for an automaton $\Auto$ is a bisimulation if
and only if   $\forall (x,y) \in \Rel$, $\forall \Cl  \in \CC$ and $\forall
a \in \Act$:
\begin{enumerate}
\item 
$\widehat{\sum}_{z \in \Cl} \trans(x,a,z) = 
\widehat{\sum}_{z \in \Cl} \trans(y,a,z)$,  equivalently $\Ea(x,\Cl) = \Ea(y,\Cl)$,
\item $\ini(x) = \ini(y)$,   equivalently $\av(x)=\av(y)$,
\item  $\val(x)=\val(y)$,  equivalently $\bv(x)=\bv(y)$,  and 
\item $AP(x) = AP(y)$ where $AP(x)$ is the set of atomic propositions
satisfied by $x$.
\end{enumerate}
\end{definition}


We define the union of two bisimulations $\Rel_1$ and $\Rel_2$ via the union
of their equivalence classes. Thus $\Rel_0 = \Rel_1 \cup \Rel_2$ is
characterized by the equivalence classes $\Clz[x]=\Clf[x]\cup\Clse[x]$ for
all $x \in \St$. With this definition the union of bisimulation relations
yields a bisimulation relation.

\begin{theorem}
Let $\Rel_1$ and $\Rel_2$ be two bisimulations for automaton $\Auto$, then
$\Rel = \Rel_1 \cup \Rel_2$ is also a bisimulation. 
\end{theorem}
\begin{proof}
  The proof is a simple extension of the proof in \cite{Buch01}, one
  needs to consider the additional condition $AP(x) = AP(y)$ of Def.
  \ref{def:bisimulation}, which is however
  straightforward. Additionally, $\Rel$ is an equivalence relation
  since it results from the union of equivalence classes.
\end{proof}

Thus, the largest bisimulation for an automaton can be defined as the union of
all bisimulations. We use the notation $x \sim y$ for $x, y \in \St$, if a
bisimulation $\Rel$ with $(x,y) \in \Rel$ exists. The bisimulation can be extended
to compare automata instead of states. This is commonly done for untimed
automata as in \cite{Miln89} but requires slight extensions if applied to the
general automata model presented here. Functions $\ini$ and $\val$ require an additional condition.
We define the union of automata
in the usual sense and bisimulation of automata by means of a bisimulation
relation  on the union.

\begin{definition} \label{def:bisim2}
Let $\Auto_1 = (\St_1, \ini_1, \trans_1,  \val_1)$ and $\Auto_2 = (\St_2, \ini_2, \trans_2,  \val_2)$ be two weighted automata defined over
the same semiring $\K$, identical alphabets $\Act$, and $\St_1\cap \St_2=\emptyset$. The union
$\Auto_1 \cup \Auto_2$ is defined as an automaton $A_0= (\St_0, \ini_0, 
\trans_0,  \val_0)$ with
\begin{itemize}
\item $\St_0 = \St_1 \cup \St_2$,
\item $\trans_0(x,a,y) = \left\{ \begin{array}{ll}
\trans_1(x,a,y) & \mbox{if } x,y \in \St_1 , \\
\trans_2(x,a,y) & \mbox{if } x,y \in \St_2 , \\
\zero & \mbox{otherwise.}
\end{array} \right.$
\item $\ini_0(x) = \ini_1(x)$ if $x \in \St_1$ and $\ini_2(x)$ for $x \in
\St_2$, and
\item  $\val_0(x) = \val_1(x)$ if $x \in \St_1$ and $\val_2(x)$ for $x \in
\St_2$.
\end{itemize}
Automata $\Auto_1$ and $\Auto_2$ are bisimulation equivalent, if a
bisimulation relation $\Rel$ exists for $\Auto_0$ and for all $\Cl \in \CC$:
\[
\widehat{\sum}_{x \in \Cl \cap \St_1} \ini(x) = \widehat{\sum}_{x \in \Cl \cap
\St_2} \ini(x) \ \mbox{ and  } \ \widehat{\sum}_{x \in \Cl \cap \St_1} \val(x)
= \widehat{\sum}_{x \in \Cl \cap \St_2} \val(x)
\]
\end{definition}

In terms of matrices, $A_0 = \Auto_1 \cup \Auto_2$ yields
\[
{\Ea}_0 = 
\left(
\begin{array}{ll} 
{\Ea}_1 & \zero \\
\zero & {\Ea}_2
\end{array}
\right) 
\mbox{ ,    }
\av_0 = \left( \av_1, \av_2 \right) 
\mbox{ ,    }
\bv_0 = 
\left(
\begin{array}{ll} 
\bv_1 \\
\bv_2
\end{array}
\right) \ .
\]

\begin{figure}[ht]
  \FIG{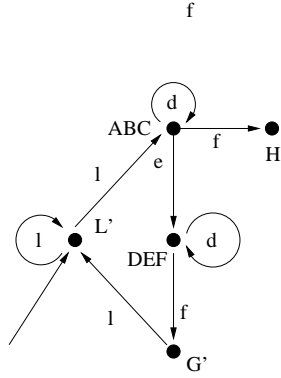} 
  \caption{Possible bisimilar model of the driving test example}
  \label{ex:modelbisim}
\end{figure}

\begin{example}\label{ex:second} Driving test, continued.   
  Fig.~\ref{ex:modelbisim} shows a model which is bisimilar to the one
  in Fig.~\ref{ex:drivingschool} provided a semiring is given and functions
  $\trans$, $\ini$ and $\val$ and sets $AP$ are appropriately defined. Let
  $AP = \{ ok, learn \}$ and $ AP(x) = \{ learn\} \forall x \in
  \{ A, B, C, D, E, F, G, G', L, L', ABC, DEF \}$ and $ AP(x) = \{
  ok\}$ for all $x \in \{ H, H' \}$. So by definition of $\ini$, $\val$,
  and $AP$, we have 3 candidates for equivalence classes $\{ H, H'
  \}$, $\{ G,G',L, L' \}$ and $\St \backslash \{G,G',L,L',H,H'\}$ to fulfill
  conditions 2-4 of the definition. By assuming $\trans(x,a,y) \not= \zero$ for all
arcs in Figs. \ref{ex:drivingschool} and \ref{ex:modelbisim} and $\zero$ otherwise, we need to partition
  $\St \backslash \{L,L',H,H'\}$ into sets $\{A,B,C,ABC\}$ and $\{D,E,F,DEF\}$ and to partition
  $\{L,L',G,G'\}$ into $\{L,L'\}$ and $\{G,G'\}$.  For the Boolean semiring,
  addition  is $\vee$ in Def. \ref{def:bisimulation}, condition 1, so it is straightforward to verify that
  this partition gives a bisimulation, i.e., $L \sim L', G \sim G', H
  \sim H', A \sim ABC, B\sim ABC, C \sim ABC, D \sim DEF, E \sim DEF,$ and $
  F \sim DEF$. If we choose the semiring $(\Rpos, +, \cdot, 0,1)$, we
  achieve the same bisimulation if we define ${\EM}_0$ for example as follows:

  \begin{center}
    \begin{tabular}[h]{c|ccccccccc|ccccc}
         & L   & A   & B   & C   & D   & E   & F   & G   & H   & L'   & ABC & DEF & G'   & H' \\ \hline
       L & ${\frac{1}{2}}_l$ & ${\frac{1}{2}}_l$ &     &     &     &     &     &     &     &      &     &     &      & \\ 
       A &     &     & ${\frac{1}{3}}_d$ &     & ${\frac{1}{6}}_e$ & ${\frac{1}{6}}_e$ &     &     & ${\frac{1}{3}}_d$ &      &     &     &      & \\
       B &     & ${\frac{1}{6}}_d$ &     & ${\frac{1}{6}}_d$ &     &     & ${\frac{1}{3}}_f$ &     & ${\frac{1}{3}}_f$ &      &     &     &      & \\
       C &     & ${\frac{1}{3}}_d$ &     &     & ${\frac{1}{9}}_e$ & ${\frac{1}{9}}_e$ & ${\frac{1}{9}}_e$ &     & ${\frac{1}{3}}_f$ &      &     &     &      & \\ 
       D &     &     &     &     &     & ${\frac{1}{2}}_d$ &     & ${\frac{1}{2}}_f$ &     &      &     &     &      & \\
       E &     &     &     &     &     &     & ${\frac{1}{2}}_d$ & ${\frac{1}{2}}_f$ &     &      &     &     &      & \\
       F &     &     &     &     &     &     & ${\frac{1}{2}}_d$ & ${\frac{1}{2}}_f$ &     &      &     &     &      & \\ 
       G & ${1}_l$   &     &     &     &     &     &     &     &     &      &     &     &      & \\ 
       H &     &     &     &     &     &     &     &     &     &      &     &     &      & \\ \hline
       L'&     &     &     &     &     &     &     &     &     & ${\frac{1}{2}}_l$  & ${\frac{1}{2}}_l$ &     &      &  \\ 
       ABC &   &     &     &     &     &     &     &     &     &      & ${\frac{1}{3}}_d$ & ${\frac{1}{3}}_e$ &      & ${\frac{1}{3}}_f$ \\ 
       DEF &   &     &     &     &     &     &     &     &     &      &     & ${\frac{1}{2}}_d$ &  ${\frac{1}{2}}_f$ & \\ 
       G'&     &     &     &     &     &     &     &     &     & ${1}_l$    &     &     &      & \\ 
       H'&     &     &     &     &     &     &     &     &     &      &     &     &      & \\ 
    \end{tabular}
    \label{tab:xxx1}
  \end{center}
The fractions give the arc weights, while the index indicates the
associated label, e.g., ${\EM}_0(L,A)={\frac{1}{2}}_l$ indicates
$\trans_0(L,l,A)=1/2$, which corresponds to a transition in the first automaton.
Matrix entries that are $\zero$ are omitted for clarity.
\end{example}
The following theorem introduces the relation between bisimulation equivalence
for weighted automata and CTL\$ formulas, which is similar to the
relation between bisimulation and CTL in untimed automata.

\begin{theorem}
If $x \sim y$, then
\begin{enumerate}
\item $x \models \Phi \ \Leftrightarrow \ y \models \Phi$ for all $\Phi$ which
are logical combinations of atomic propositions,
\item $x \models [a]_{\bowtie p}.\Phi  \ \Leftrightarrow \ y \models [a]_{\bowtie
p}.\Phi$,
\item $x \models \ufg \ \Leftrightarrow \ y \models \ufg$ and 
\item $x \models \aufg \ \Leftrightarrow \ y \models \aufg$.
\end{enumerate}
where $\Phi_1$ and $\Phi_2$ are CTL\$ formulas.
\end{theorem}
\begin{proof}
1. holds since $AP(x) = AP(y)$ for $x \sim y$ such that also all logical
   combinations of atomic propositions yield identical results.

2. is proved inductively by assuming that for $x \sim y$: $x \models
   \Phi \Leftrightarrow y \models \Phi$. Then   $x \models [a]_{\bowtie p}.\Phi  \ \Leftrightarrow \ y \models [a]_{\bowtie
p}.\Phi$ since $\Ea(x,\Cl) = \Ea(y,\Cl)$. Initially we know that $AP(z)$ is the same for
   all $z \in \Cl$ such that the relation holds for all $\Phi$ which
   are logical combinations of atomic propositions. By induction the
   relation also holds for $\Phi$ containing an arbitrary number of
   constructs of the form $[a]_{\bowtie p} \Phi$. For more general
   formulas we combine the induction used in this step with the
   induction presented for 3. and 4. below. 

3. and 4. have to be proved inductively over the number of occurrences
of $\ufg$ and $\aufg$ in the formula and over the length $t$ of
the required paths. First assume for $x \sim y$:
\[
x \models \Phi_1 \ \Leftrightarrow y \models \Phi_1 \mbox{ and }
x \models \Phi_2 \ \Leftrightarrow y \models \Phi_2
\]
which is proved for formulas $\Phi_1$ and $\Phi_2$ that do not contain
$\ufg$ or $\aufg$. Now we prove $x \models \ufg \Leftrightarrow  y
\models \ufg$ inductively over $t$. For $t=0$ we have:
\[
\av(x) \bv(x) = \av(y) \bv(y) 
\] 
such that the formula holds for $x\sim y$. 
Define
for $\Cl \in \CCm$: $\xi(\Cl) = \bv(z)$ for some (all) $z \in \Cl$ and
let $\delta(\Cl, \Phi) = \one$ if some (all) $z \in \Cl$: $z \models \Phi$
and $\zero$ otherwise.
For $t = 1$, the following relation holds for  $x\sim y$ and all $\Cl \in \CCm$:
\[
\begin{array}{ll}
\av(x)\widehat{\sum}_{z \in \Cl} \Eoyty(x,z)\bv[\Phi_2](z) & \\ \\
\av(x)\Eoyty(x,\Cl)\xi(\Cl)\delta(\Cl,\Phi_2)  & = \\ \\
\av(y)\Eoyty(y,\Cl)\xi(\Cl)\delta(\Cl,\Phi_2)  & = \\ \\
\av(y)\widehat{\sum}_{z \in \Cl} \Eoyty(y,z)\bv[\Phi_2](z)
\end{array}
\]
such that the required property is given for $t=1$.
Let $\bv'_C(i) = \widehat{\sum}_{z \in \Cl} \Eoyty(i,z)\bv[\Phi_2](z)$
for $i \in \St$. So, the aforegoing argumentation ensures that
$\bv'_C(x)=\bv'_C(y)$ if $x\sim y$.

For the induction step, we
assume that the relation has been proved for $t \ge 1$ and we show that it
holds for $t+1$. To simplify the notation, let 
$\PM = \Eoytn$, $\QM = \Eoyty$, and 
$\RM_t = \left(\widehat{\sum}_{k = 0}^{t}(\Eoytn)^k\right) \Eoyty$.
We have to prove that 
\[
\av(x)\widehat{\sum}_{z \in \Cl}\RM_t(x,z) \bv[\Phi_2](z)=\av(y)\widehat{\sum}_{z \in \Cl}\RM_t(y,z) \bv[\Phi_2](z) \ .
\]
Since $\av(x)=\av(y)$ for $x\sim y$, we only need to show that
\[
\widehat{\sum}_{z \in \Cl}\RM_t(x,z) \bv[\Phi_2](z)=\widehat{\sum}_{z \in \Cl}\RM_t(y,z) \bv[\Phi_2](z) \ .
\]
Starting from the left side, we obtain by the induction assumption, 
\[
\begin{array}{ll}
\widehat{\sum}_{z \in \Cl}\RM_t(x,z) \bv[\Phi_2](z) & = \\
\widehat{\sum}_{z \in \Cl}\RM_{t-1}(x,z) \bv[\Phi_2](z) \hp 
\widehat{\sum}_{z \in \Cl}\left(\PM^t\QM \right)(x,z) \bv[\Phi_2](z) &
= \\
\widehat{\sum}_{z \in \Cl}\RM_{t-1}(y,z) \bv[\Phi_2](z) \hp 
\widehat{\sum}_{z \in \Cl}\left(\PM^t\QM \right)(x,z) \bv[\Phi_2](z) &
= \\
\widehat{\sum}_{z \in \Cl}\RM_{t-1}(y,z) \bv[\Phi_2](z) \hp 
\idn_x \PM^t \bv'_C &
\end{array}
\]
At this point, we are done if $\idn_x \PM^t = \idn_y \PM^t$. Obviously, $x \sim y$ implies only $\PM(x,\Cl') = \PM(y,\Cl')$ for all $x,
y \in \Cl$ and all $\Cl, \Cl' \in \CCm$. However, we can show by induction that
%
$\PM(x,\Cl') = \PM(y,\Cl') = \psi_1({\Cl,\Cl'})$ for all $x,
y \in \Cl$ and all $\Cl, \Cl' \in \CCm$ implies $\PM^k(x,\Cl) \PM^k(y,\Cl)= \psi_k({\Cl,\Cl'}) $ for $k>0$. By definition, the
statement is true for $k=1$. So for an inductive argument, we can assume
that the result holds for $k-1$, then we have for an arbitrary $\Cl
\in \CCm$ and all $x, y \in \Cl$:
\[
\begin{array}{llll}
\PM^k(x,\Cl') & = & \widehat{\sum}_{z \in \St} \PM(x,z)  \PM^{k-1}(z,\Cl') &
= \\ \\
\widehat{\sum }_{\Cl'' \in \CCm} \widehat{\sum}_{z \in \Cl''} \PM(x,z)
\PM^{k-1}(z,\Cl') & = &
\widehat{\sum }_{\Cl'' \in \CCm} \widehat{\sum}_{z \in \Cl''} \PM(x,z)
\psi_{k-1}(\Cl'',\Cl') & = \\ \\
\widehat{\sum }_{\Cl'' \in \CCm} \psi_1(\Cl,\Cl'') 
\psi_{k-1}(\Cl'',\Cl') & = &
\widehat{\sum }_{\Cl'' \in \CCm} \widehat{\sum}_{z \in \Cl''} \PM(y,z)
 \psi_{k-1}(\Cl'',\Cl') & = \\ \\
\widehat{\sum }_{\Cl'' \in \CCm} \widehat{\sum}_{z \in \Cl''} \PM(y,z)
\PM^{k-1}(z,\Cl') & = &
 \widehat{\sum}_{z \in \St} \PM(y,z) \PM^{k-1}(z,\Cl') & = \\ \\
\PM^k(y,\Cl')
\end{array}
\]
So in summary, we obtain 
\[
\begin{array}{ll}
\widehat{\sum}_{z \in \Cl}\RM_t(x,z) \bv[\Phi_2](z) & = \\
\widehat{\sum}_{z \in \Cl}\RM_{t-1}(y,z) \bv[\Phi_2](z) \hp 
\idn_x \PM^t \bv'_C & = \\
\widehat{\sum}_{z \in \Cl}\RM_{t-1}(y,z) \bv[\Phi_2](z) \hp 
\idn_y \PM^t \bv'_C & = \\
 \widehat{\sum}_{z \in \Cl}\RM_t(y,z) \bv[\Phi_2](z) \\
\end{array}
\]
and the induction step is complete.
This finishes considerations of $\ufg$.

For $\aufg$, the above line of argumentation can be used completely analogously to prove
\[
\begin{array}{c}
\av(x) \idn_x (\widehat{\sum}_{k=0}^{t-1}(\Eoytn)^k)  \Eontn \id  =  \zero  \\
\Leftrightarrow \\
\av(y)  \idn_y  (\widehat{\sum}_{k=0}^{t-1}(\Eoytn)^k)  \Eontn \id  = \zero
\end{array}
\]
and 
\[
\av(x)  \idn_x (\Eoytn)^t \id = \zero \ \Leftrightarrow \ 
\av(y)  \idn_y(\Eoytn)^t\ \id = \zero
\]
which proves $x \models \aufg \ \Leftrightarrow \ y \models \aufg$. We
omit the details, since they provide no further insight. 

Finally, to prove $\ufg$ and $\aufg$ for general $\Phi_1$ and $\Phi_2$, we again
use induction, namely over the number of occurrences of $\ufg$ or $\aufg$ in a
formula. Note that in the aforegoing 
argumentation, we did not use any other assumption for $\Phi_1$ and
$\Phi_2$ than that $x \models \Phi_1 \ \Leftrightarrow y \models \Phi_1 \mbox{ and }
x \models \Phi_2 \ \Leftrightarrow y \models \Phi_2$. Since this
assumption holds here again by the induction assumption we can simply
repeat the argumentation for $\ufg$ and $\aufg$ above for the induction step.
\end{proof} 

The above theorem shows that 
one cannot distinguish between bisimilar
states or automata by model checking CTL\$ formulas.  Thus, an automaton can be first reduced according
to bisimulation equivalence to gain efficiency in subsequent model
checking algorithms. For this purpose, first relation $\sim$ is
computed, which can be done by a partition refinement algorithm, and
then each equivalence class of $\sim$ is substituted by a single
state, which yields an aggregated automaton \cite{Buch01}. Afterwards,
formulas are checked with the aggregated instead of the original
automaton. In \cite{BuKe01}, it is shown that bisimulation is a
congruence according to the composition operators of the process
algebra GPA, which allows compositional analysis by interleaving
reduction of components due to bisimulation equivalence and
composition of components. In this way, a reduced automaton is
generated to which model checking is applied.

\section{Examples of automata with specific semirings}
\label{sec:examples}

We present six examples in the following subsections. Two of the examples
describe known  types of automata which are presented in the proposed
framework. In these cases we show that CTL\$ model checking is related to
logics presented specifically for these automata types. Afterwards we present
new approaches for model checking. 

\subsection{Untimed automata}

Untimed automata are defined over the semiring $(\B, \vee, \wedge, 0, 1)$. For
these automata $\ini(x)=1$ for initial states and $\ini(x)=0$ for the
remaining states. Similarly, $\val(x)=1$ for terminating states and $0$ for
the remaining states. $\trans(x,a,y)=1$ describes the existence of an
$a$-labeled transition between $x$ and $y$. The Boolean semiring is ordered
($0 < 1$), the order is preserved by the operations and $0$ is the
infimum. Therefore the conditions we proposed for model checking are
observed. In the Boolean case all paths have the same weights, namely
$1$.

For untimed automata CTL is a logic which is often used for model checking. We
now show how the path formulas of CTL can be expressed by CTL\$. State formulas
defined via atomic propositions are obviously identical in both cases. \bigskip
\begin{center}
\begin{tabular}{|c|c|}
\hline CTL & CTL\$ \\ \hline
$EX \Phi$ & $true \ U_{> 0}^1 \ \Phi$ \\
$AX \Phi$ & $true \ U_{> 0}^1 \ \Phi$ \\
$A[\Phi_1 \ U \ \Phi_2]$ & $\Phi_1 \ AU_{> 0}^{\infty} \ \Phi_2$ \\
$E[\Phi_1 \ U \ \Phi_2]$ & $\Phi_1 \ U_{> 0}^{\infty} \ \Phi_2$ \\
$AF \ \Phi$ & $true \ AU_{> 0}^{\infty} \ \Phi$ \\
$EF \ \Phi$ & $true \ U_{> 0}^{\infty} \ \Phi$ \\ 
$AG \ \Phi$ & $\neg \ (true \ U_{> 0}^{\infty} \ \neg \Phi_1)$ \\
$EG \ \Phi$ & $\neg (true \ AU_{> 0}^{\infty} \ \neg \Phi_1)$ \\\hline  
\end{tabular}
\end{center}\bigskip

For the detailed description of the CTL-formulas see \cite{ClES86}. It is easy
to show that for the Boolean case the model checking algorithms proposed above
all have a finite runtime because for $\eufg$, $\aufg$ and $\ufg$ are
identical for all $t \ge n$. The reason for this behavior is that for an
automaton with $n$ states
between two states a path of length $\le n$ or no path exists and since
additionally all paths have the same weights and addition is
idempotent, it is sufficient to consider paths
up to length $n$ if no longer paths have been defined explicitly via
concatenation of $[a]$ in the formulas. Consequently, the following
relation holds for the Boolean semiring.
\[
\Noytn = \widehat{\sum}_{k=0}^{\infty} (\Eoytn)^k = \widehat{\sum}_{k=0}^{n} (\Eoytn)^k
\]

For the representation of CTL formulas using CTL\$, paths of arbitrary length
are considered. However, by considering paths of finite length and assuming
that each transition of the automaton has a duration of one time unit, real
time properties can be proved by CTL\$ model checking. In this case CTL\$ can be
used to mimic formulas of the real time logic RTCTL \cite{EMSS92}.

\begin{example}
We consider the driving test example shown in Fig.
\ref{ex:drivingschool} over the Boolean semiring. In this case, each arc in
the graph describes a transition with weight $1$. Since for the Boolean case
bisimilar automata cannot be distinguished by CTL\$, model checking can be performed
using the aggregated automaton shown in Fig. \ref{ex:modelbisim}. 

Since $L'$ is the only initial state of the automaton, we have to prove whether
a formula holds for $L'$. Formula $(true \ U_{> 0}^{\infty} \ ok)$ states that it
is possible to pass the driving examination in an arbitrary number of
steps. This formula is obviously satisfied by $L'$. The shortest path
satisfying the formula start in $L'$ passes $ABC$ and then enters $H'$. Thus,
also the formula $(true \ U_{> 0}^{t} \ ok)$ is satisfied by $L'$ for all $t >
1$ which means that the driving examination can be passed in 2 steps. The
formula $(true \ AU_{> 0}^{\infty} \ ok)$ states that the examination is always
passed. This formula is not satisfied by $L'$ because paths of infinite length
exist which do not reach $H'$. Consequently, also formula $(true \ AU_{> 0}^{t}
\ ok)$ does not hold for $L'$.

In the Boolean semiring it is not possible to determine more detailed results
about reaching state $H'$. We can only state that a path exists which reaches
$H'$ and that not all paths reach $H'$. 
CTL\$ allows us to
derive results about the length of the path reaching $H'$ but not about the
quantification of paths because the $U$ operator equals $EU$ in the Boolean
semiring. This is different in the other semirings, we consider in the
subsequent paragraphs. 
\end{example}

\subsection{Probabilistic automata}
\label{sec:probauto}

Probabilistic automata are defined over the semiring $(\Rpos, +,\cdot, 
0, 1)$ with the additional restrictions
\[
\begin{array}{c}
\sum_{x \in \St} \ini(x) = 1 \ , \\ \\
\sum_{a \in \Act} \sum_{y \in \St} \trans(x,a,y) = 1 \mbox{ for all } x \in
\St \ .
\end{array}
\]
A probability distribution is defined as the initial distribution and the sum
of transition probabilities leaving a state is $1$. 
These restrictions define
a generative probabilistic model in the sense of
\cite{GSST90} because the automaton decides probabilistically which
transition occurs next. Additionally, the automata model is similar to the
model presented in \cite{HaJo94} with additional possibility of labeling
transitions. Probabilistic automata are ordered, the order is preserved by the
operations and $0$ is the infimum which implies that model checking can be
applied for this automata type.

For probabilistic automata the logic PCTL has been proposed in
\cite{HaJo94}. This logic contains, apart from  state propositions and logical
combinations of state propositions, the path quantifier $\ufg$ with a similar
semantics as in CTL\$. For finite $t$, the following relation between
the path formulas $U$ and $AU$ holds in probabilistic systems.
\[
\begin{array}{c}
x \models \Phi_1 \ U_{\ge 1}^t \ \Phi_2 \ \Leftrightarrow 
x \models \ \Phi_1 \ AU_{\ge 1}^t \ \Phi_2 \\ \\
\end{array}
\]
The above relation does not necessarily hold for $t = \infty$ as shown
in the example below.

Now we consider CTL\$ model checking for probabilistic automata. Interesting
are the formulas  $\Phi_1 \
U_{\bowtie p}^{\infty} \ \Phi_2$. For the remaining cases the algorithms
presented in section \ref{sec:model-checking} can be used because they require 
in these cases a finite number of steps. For the formulas with $t = \infty$, matrix $\Noytn$ has to be computed first which can be done as
shown in the following theorem. 

\begin{theorem}
\label{path-theorem}
State $x \in \St$ satisfies formula $\Phi_1 \ U_{\bowtie p}^{\infty} \ \Phi_2$
if $\EM[\Phi_1\wedge \neg \Phi_2 ,\Phi_2]$ is a substochastic
matrix  without a stochastic submatrix and
\[
\begin{array}{ll}
\av(x) (\widetilde{\sum}_{k = 0}^{\infty}
(\Eoytn)^k) \Eoyty \bv[\Phi_2]
& = \\ \\
\av(x) \left( \Id - \Eoytn\right)^{-1} \Eoyty \bv[\Phi_2] 
& \bowtie p
\end{array}
\]
\end{theorem} 
\begin{proof}
The matrix representation of the formula has already been introduced. The relation
\[
\sum_{k=0}^{\infty} \left(\Id - \Eoytn \right)^k 
= \left(\Id - \Eoytn \right)^{-1}
\]
is well known for absorbing Markov chains \cite{KeSn76} under the conditions
stated in the theorem.
\end{proof}

The theorem contains a method to decide for which states $\Phi_1 \ U_{\bowtie
p}^{\infty} \ \Phi_2$ holds. If $\Eoytn$ contains a
stochastic submatrix, then there exists a subset of states where $\Phi_1$ but
not $\Phi_2$ holds and this subset of states forms a trap according to
the formula, i.e., the automaton
can never leave the subset after entering it. Obviously $\Phi_1 \ U_{\bowtie
p}^{\infty} \ \Phi_2$ cannot be satisfied in these states and each path
entering the subset does not count when path weights are summed. Thus, for
subsets of states forming a stochastic submatrix, the rows in 
$\Eoytn$ might be
set to ${\bf 0}$ to compute the result. After this modification the inverse  
matrix exists and the
set of states satisfying $\Phi_1 \ U_{\bowtie p}^{\infty} \ \Phi_2$ can be
computed in finitely many steps.

\begin{example}
For the probabilistic case, we consider the driving test example with probabilities
given in Examp. \ref{ex:second} where also a bisimilar automaton with less
states is presented. We can check the smaller aggregated
automaton and consider the formula ($true \ U_{\ge p}^{\infty} \ ok$) which is
true if the test is passed with probability of at least $p$ in an arbitrary
number of steps. State $L$ satisfies the formula if 
\[
\av(L') \sum_{k=0}^{\infty} \left( {\bf T}[true, ok] \right)^k \cdot \left(
\bv[ok]\right) \ge p
\] 
holds. Using the ordering of states $(L',ABC,DEF,G',H')$ as given in \ref{ex:second} we obtain the
following matrices and vectors.
\[
\av = \left(
1, 0, 0, 0, 0 \right), \ \
\bv[ok] = \left( 0, 0, 0, 0, 1 \right)^T
\]
\[
{\bf T}[true, ok] = \left( \begin{array}{ccccc}
1/2 & 1/2 & 0   & 0   & 0 \\
 0  & 1/3 & 1/3 & 0   & 1/3 \\
 0  &  0  & 1/2 & 1/2 & 0  \\
 1  &  0  & 0   & 0   & 0 \\
 0  &  0  & 0   & 0   & 0 
\end{array} \right)
\]
\[
\left({\bf I} - {\bf T}[true, ok]\right)^{-1} = \left( \begin{array}{ccccc}
 4  &  3  & 2   & 1   & 1 \\
 2  &  3  & 2   & 1   & 1 \\
 4  &  3  & 4   & 2   & 1  \\
 4  &  3  & 2   & 2   & 1 \\
 0  &  0  & 0   & 0   & 1 
\end{array} \right)
\]
and 
\[
\av \left({\bf I} - {\bf T}[true,
  ok]\right)^{-1}\left(\bv[ok]\right)^{T} \left(1, 1, 1, 1, 1\right)  1 
\]
which implies that the formula is observed for all $p \le 1$. This means that
after an arbitrary number of steps the driving test will be passed with
probability $1$. However, $\neg L' \models true \ AU_{\ge p}^{\infty} \ ok$
for all $p$. The example nicely shows that probability $1$ does not mean that
the result holds for all paths. This result is, of course, well known from
probability theory. 
\end{example}

\subsection{Max/plus automata}

Max/plus automata are defined over the completed semiring $(\Rpos \cup
\{-\infty,\infty \}, \max, +, -\infty, 0)$ with the computation $a \hp -\infty
= \max(a, -\infty) =a$ and $a \hm 0 = a + 0 = a$. The weights of a path
in max/plus correspond to the sum of weights of each transition on the
path because multiplication is represented by the usual addition. If
we consider several paths, then the maximum operator computes the
weights of the path with the highest weights. Max/plus automata can be
applied for various analysis purposes including the analysis of real
time systems or communications networks and became very popular in the
recent years. The max/plus semiring is ordered according to the usual
ordering $a \le b \ \Leftrightarrow \ \max(a,b)=b$. Furthermore, the
order is preserved by the operations and $\zero$, in this case
$-\infty$ is the infimum of the semiring. In the definition of the
transition function $T$, $-\infty$ is used to denote that an arc does
not exist, which is the common usage of element $\zero$. We can
directly apply our model checking approach. 

Computation of the matrix $\Noytn$ requires the analysis of cycles in
the matrix $\Eoytn$. $x_1,\ldots, x_K$ is a cycle if
$\Eoytn(x_{k},x_{k+1})\neq\zero$ ($1\le k < K$) and $x_1 = x_K$. The
cycle has a positive weight if $\widehat{\prod}_{k=1}^{K-1}
\Eoytn(x_k,x_{k+1})>0$. It is well known that all cycles can be
generated by composing minimal cycles and minimal cycles can be
computed using some standard algorithms from graph theory.  Element
$\Noytn(x,y)=\infty$, if a minimal cycle with a positive weight that
contains arc $(x,y)$ exists. The remaining elements in matrix
$\Noytn$, which are not $\infty$ can be computed from
$\widehat{\sum}_{k=0}^n (\Eoytn)^k$.

\begin{example}
   For this semiring, we need a different selection of $T(x,a,y)$ to achieve 
bisimilar automata in Figs. \ref{ex:drivingschool} and \ref{ex:modelbisim}, since the maximal values of outgoing
arcs of bisimilar states leading to the same class of states need to
be equal. We select the following values which may be interpreted as
distances the student has to drive or a quantification of the amount
of stress he/she has to suffer. Matrix elements that are $\zero$ and
transition labels are omitted for clarity.

  \begin{center}
    \begin{tabular}[h]{c|ccccccccc|ccccc}
         & L   & A   & B   & C   & D   & E   & F   & G   & H   & L'   & ABC & DEF & G'   & H' \\ \hline
       L & 1   & 2   &     &     &     &     &     &     &     &      &     &     &      & \\ 
       A &     &     &   3 &     &   6 &   9 &     &     &   3 &      &     &     &      & \\
       B &     & 1   &     & 3   &     &     & 9   &     &   3 &      &     &     &      & \\
       C &     &   3 &     &     &   9 &   4 &  7  &     &   3 &      &     &     &      & \\ 
       D &     &     &     &     &     &   2 &     & 1   &     &      &     &     &      & \\
       E &     &     &     &     &     &     &   2 & 1   &     &      &     &     &      & \\
       F &     &     &     &     &     &     &   2 & 1   &     &      &     &     &      & \\ 
       G & 1   &     &     &     &     &     &     &     &     &      &     &     &      & \\ 
       H &     &     &     &     &     &     &     &     &     &      &     &     &      & \\ \hline 
       L'&     &     &     &     &     &     &     &     &     &   1  &   2 &     &      &  \\ 
       ABC &   &     &     &     &     &     &     &     &     &      & 3   & 9   &      & 3 \\ 
       DEF &   &     &     &     &     &     &     &     &     &      &     & 2   &  1   & \\ 
       G'&     &     &     &     &     &     &     &     &     & 1    &     &     &      & \\ 
       H'&     &     &     &     &     &     &     &     &     &      &     &     &      & \\ 
    \end{tabular}
    \label{tab:xxx2}
  \end{center} For instance, $A \sim ABC$ since $T(A,d,B) = 3 = T(ABC,d,ABC)$, $max(T(A,e,D),T(A,e,E))=9=T(ABC,e,DEF)$, and
       $T(A,f,H)=3=T(ABC,f,H')$ and further conditions of
       Def. \ref{def:bisimulation} with respect to $\ini$, $\val$ and
       $AP$ hold as well.

  In this semiring, CTL\$ considers the most costly (or stressful)
  ways to a driver's license exist.  E.g. one can compute by the
  algorithm given in Sec. \ref{sec:model-checking} that $L \models
  true U^{9}_{\ge 21} ok$ holds due to path $\pi$ through states
  $L,A,E,G,L,A,B,H$ with $ce(\pi) = 21$.  For model checking we can use the bisimilar
  automaton given above, which contains less states and less arcs. So
  we check $L'\models true U^{9}_{\ge 21} ok$ which holds due to path
  $\pi'$ through states $L',ABC,EFG,G',L',ABC,ABC,H$. Both paths are
  of same length and have the same weights.
\end{example}

\subsection{Min/plus automata}

The min/plus approach is very similar to the max/plus approach. 
It is applied if one is interested in minimal weights instead of maximal
weights. It is defined on the semiring $(\Rpos\cup \{\infty \},min,+,\infty,0)$ with an
inverse order, i.e.  addition becomes minimum $ x\hp y = min(x,y)$,
multiplication becomes addition $ x\hm y = x + y$, and $x$, $y$ are
ordered $x \ge y$ iff $x = min(x,y)$. The semiring preserves the order
and $\zero=\infty$ is the infimum. Working with an inverse order is
formally correct but rather contrary to intuition. Note that the
inverse order of min/plus is the reason to use the notion of infimum
and supremum rather than minimum and maximum in this paper. This avoids
reformulation of the algorithms.  Model checking algorithms can be
applied analogously as for max/plus automata. 

Computation of the matrix $\Noytn$ is easier for min/plus than for
max/plus. The reason is that minimal weights count. Since a cycle
cannot reduce the weight of a path we have, as in the Boolean
semiring 
\[
\Noytn = \widehat{\sum}_{k=0}^{\infty} (\Eoytn)^k = \widehat{\sum}_{k=0}^{n} (\Eoytn)^k
\]


\MSG{Hier Beispiel fuer minimale Kosten auf bisimularem Automaten durchgehen,
  \begin{itemize}
  \item zuerst transitions funktion mit minimalen werten fuer automaten definieren
  \item daran modelchecking fuer EU und U durchfuehren
  \item Werte koennen als Stressfaktor interpretiert werden (``subjektiv'')
  \item Werte koennen auch als Kosten je gefahrene Strecke gezaehlt werden
  \end{itemize}
}

\subsection{Max/min automata}

The semiring $(\Rpos\cup \{\infty \},max,min,0,\infty)$ is useful to
identify paths with respect to bottlenecks, since the weight of a path
gives the minimum value observed through all of its arcs. We consider
a communication network as an example. A weighted automaton models
the network by using nodes for hubs and arcs for links between
hubs. Each hub shows a certain utilization and each link has a
bandwidth as a non-negative real number assigned to it. If there is no
link between two nodes, we assume an arc with weight $0$. In order to
establish a point to point communication, we are interested in the
existence of a connection between two nodes $x_{start}$ and $x_{end}$ that has a minimum bandwidth $\mu$, uses less
than $\lambda$ intermediate nodes and the employed nodes should have a
utilization less than $\gamma$ to avoid saturated or overloaded nodes.

To express this in CTL\$, we first define atomic propositions
$\Phi_1$ and $\Phi_2$ as follows.
A node $x \models \Phi_1$ if and only if its utilization is less than
$\gamma$.
A node $x \models \Phi_2$ if and only if it is node $x_{end}$.
The following formula describes the property we are interested in
$ \Phi = \Phi_1 \ U_{\geq p}^t \ \Phi_2$ with $p= \mu$, $t = \lambda$.
Model checking the automata by the algorithm for $\Phi_1 \ U_{\bowtie
  p}^t \ \Phi_2$ given in Section \ref{sec:model-checking}  provides us with
information whether $x_{starts} \models \Phi$ or not.
Note that the semiring is ordered, the operations preserve the order
and $\zero=0$ is the infimum.

Computation of the matrix $\Noytn$ is easy in the max/min semiring
because the weight of a path is determined by minimum weight of an arc
on this path such that cycles cannot increase the weight of a path and
$\Noytn$ can be computed by finite summation like for the Boolean
semiring or the min/plus semiring.

\subsection{The expectation semiring}

In this subsection, we consider a semiring which is more complex than
the previous and outline how our modelchecking approach can be
extended to analyze also this system. However, the extension requires
some additional steps. The proposed semiring is motivated by a semiring
given in \cite{Eisn01} and 
allows the simultaneous computation of path probabilities and expected
values of a set of paths. A value in the expectation semiring consists
of two components $(p,v)$ with $p, v \in \Rpos$. The
operations are defined as
\[
(p_1,v_1) \hm (p_2,v_2) = (p_1\cdot p_2, v_1 + v_2) \mbox{ and } 
(p_1,v_1) \hp (p_2,v_2) = (p_1 + p_2, (p_1\cdot v_1 + p_2 \cdot
v_2)/(p_1 + p_2))
\]
where $0 / 0 = 0$. We have $\zero = (0, 0)$ and $\one = (1,
0)$. Furthermore, we define the ordering $\ge$ with $(p_1,v_1) \ge
(p_2,v_2)$ if $p_1 \ge p_2$ and $v_1 \le v_2$. Observe that this
defines only a partial order since elements exist where neither
$(p_1,v_1) \ge (p_2,v_2)$ nor $(p_1, v_1) \le (p_2, v_2)$ holds. The
semiring is commutative because multiplication is commutative, it is
not idempotent, and it is also not order preserving. 

Assume that we have an automaton over the expectation semiring where
all transitions are labeled with a single label which will be
suppressed in the sequel. As in probabilistic automata, let the sum of the first components $p_i$ of the
weights $(p_i,v_i)$ of transitions $i=1,2,\dots$ that leave a state be smaller or equal to
$1$. Thus, the first values form a probability distribution of
choosing a successor state, the second components might be interpreted
as the costs of a transition. Assume that $\ini(s)=\one = (1,0)$ for one
state, the initial state. Assume further that a predicate $\Phi_2$ for
one state $s'$ with $\val(s') = \one$. Formula $true \ U_{\bowtie (p,v)}^t
\Phi_2$ holds if the probability of reaching the final state from the
initial state in at most $t$ steps is at least $p$ and the expected
costs are smaller or equal $v$. Similarly $\ufg$ holds if only nodes
are touched where $\Phi_1$ holds on the way from the initial state to
the final state, the remaining conditions are as in the previous
case. For finite $t$ the formulas can be checked with the proposed
algorithms  after some modifications that change all parts which are
based on the order preserving property of the semiring.

For this specific semiring, even results for $t = \infty$ may be
checked but this requires some tools from the analysis of Markov
processes which are beyond the scope of this paper.

The introduction of this rather unconventional semiring shows that the
proposed method modelchecking approach can be extended to a very large class of
models by using sophisticated semirings. However, if these semirings
do not fall into the basic class defined at the beginning of section
\ref{sec:model-checking}, model checking algorithms have to be adjusted specifically to
their properties. 

\section{Conclusions}
\label{sec:conclusion}

We present a general approach for model checking weighted automata 
which covers classical types of automata like untimed
or probabilistic automata as well as new types like max/plus and
min/plus automata.  The key idea is that transitions weights can be
taken of an arbitrary ordered semiring which is an algebraic structure
of very modest requirements. We present a modal logic CTL\$ for this
class of models that is build on top of CTL and allows to specify
paths with respect to their length and weights.  This yields a generic
approach where new, different semirings automatically profit from
algorithms and results derived for the general case, e.g. we present a
bisimulation for CTL\$ that is subsequently used to modelcheck an
example under various weight assignments of different semirings.  So far
we presented analysis algorithms based on graphs assuming an explicit
representation of states.  This was for clarity and to limit the scope
of the paper. Clearly, large state spaces are better treated by a
symbolic representation. In the special case of the boolean semiring,
binary decision diagrams (BDDs) and corresponding algorithms
\cite{ClGP99} are sufficient. However the general case requires the
treatment of numerical values, such that corresponding extensions of
BDDs like multi-terminal BDDs as in \cite{ClWi96} for instance are more
appropriate. Furthermore, compositional representations as in \cite{BuKe01}
and compositional model checking are interesting candidates for
modelchecking CTL\$. 

Advantages of the approach presented here are foreseen for building
analysis tools and to allow for model checking in different
application areas like realtime scheduling and logistic networks. The
latter is in the focus of a large DFG-funded collaborative research
centre (SFB 559), with significant interest in modelchecking.  The
development of tools profits from our approach since it nicely matches
an object oriented design, where model checkers of specific semirings
can inherit functionality from an implementation of the general case.
We currently work to integrate this approach into an existing CTL
modelchecker within the APNN toolbox \cite{BaBK98}.

\bibliographystyle{plain}
\bibliography{/home/buchholz/TeX/misc/paperbib}
\end{document}